\documentclass{epl}
\usepackage{graphicx}
\usepackage{amsmath}
\usepackage{amsfonts}
\usepackage{amssymb}

\newcommand{\be}{\begin{equation}}
\newcommand{\ee}{  \end{equation}}
\newcommand{\ba}{\begin{eqnarray}}
\newcommand{\ea}{  \end{eqnarray}}

\title{Antiresonances as precursors of decoherence}

\author{L. E. F. Foa Torres\inst{1} \and H. M. Pastawski\inst{2} \and E. Medina\inst{3}}
\institute{
  \inst{1} International Center for Theoretical Physics, Strada Costiera 11, 34014 Trieste, Italy\\
  \inst{2} Facultad de Matem\'{a}tica, Astronom\'{\i}a y F\'{\i}sica, Universidad Nacional de C\'{o}rdoba, Ciudad Universitaria, 5000 C\'{o}rdoba, Argentina\\
  \inst{3} Centro de F\'{\i}sica, Instituto Venezolano de Investigaciones Cient\'{\i}ficas, Apartado 21827, Caracas 1020A, Venezuela
}

\begin{document}

\maketitle

\begin{abstract}
We show that, in presence of a complex spectrum, antiresonances act as a
precursor for dephasing enabling the crossover to a fully decoherent transport
even within a unitary Hamiltonian description. This general scenario is
illustrated here by focusing on a quantum dot coupled to a chaotic cavity
containing a finite, but large, number of states using a Hamiltonian
formulation. For weak coupling to a chaotic cavity with a sufficiently dense spectrum, the ensuing complex structure of resonances and antiresonances leads to phase randomization under coarse graining in energy. Such phase instabilities and coarse graining  are the ingredients
for a mechanism producing decoherence and thus irreversibility.
For the present simple model one finds a conductance that coincides with the one obtained by adding a ficticious voltage probe within the Landauer-B$\ddot{\rm u}$ttiker picture.
This sheds new light on how the microscopic mechanisms that produce phase
fluctuations induce decoherence.

\end{abstract}

In the last decade the quantum-classical transition has been object of intense
study leading to a substantial progress in its
comprehension \cite{cit-RMP-Zurek}. An essential ingredient is the fact that the properties of a given system, though simple, will be influenced by a hierarchy of interactions with the
rest of the universe (\textit{the environment}). However weak, such
interactions lead to the degradation of the ubiquitous interference phenomena
characteristic of a quantum system, i.e., decoherence. Current trends in
technology focus on the tailoring of such interference phenomena to achieve
different goals. These range from the control of electronic currents at the
nanoscale in semiconductor and molecular devices \cite{cit-review-MOLelect,PecchiaRepProgPhys2004} to
the flow of quantum information \cite{cite-Zeilinger} encoded in the phase of a
quantum state. The crossover between the coherent and decoherent dynamics
is manifest in the behavior of the quantum phase strikingly
exhibited \cite{cit--Webb-Washburn} by weak localization phenomena and the
Aharanov-Bohm effect. Hence, the understanding and control of the effects of
the coupling to the environment on the quantum phase constitute a central
problem for both, fundamental physics and practical applications.

\begin{figure}[ptb]
\vspace{2.5cm}
\onefigure[width=8cm]{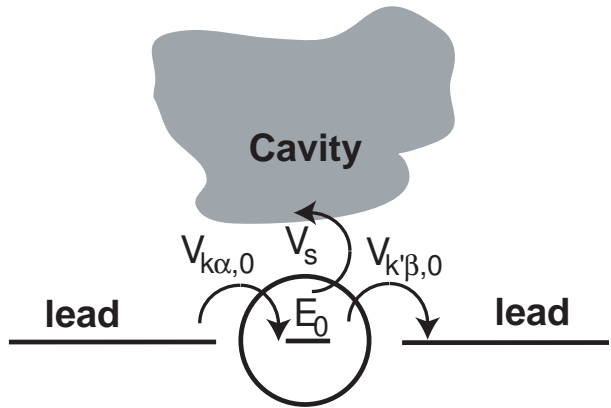}
\caption{Schematic representation of the model system considered in the text: A quantum dot
connected to left and right leads and to a chaotic cavity whose effect is the
focus of this work.}%
\label{fig-model}%
\end{figure}

The most obvious source of decoherence is the creation of entangled
system-environment states induced by complex many-body interactions. However,
recent results on the Loschmidt Echo in chaotic systems \cite{cit--LoschmidtEcho} have suggested that complexity is a natural road to decoherence even in
a one body problem. Indeed, once the system is complex enough there is little
chance to sustain a controllable interference experiment. In this paper, we
will explore the notion that, in presence of a complex spectrum,
antiresonances are a precursor for dephasing and result in decoherent
transport even within a fully unitary Hamiltonian description. This is
illustrated by considering a toy model for a quantum dot tunneling device
coupled to a chaotic cavity containing a \textit{large, but finite}, number of
states in the energy range of interest. A possible arrangement is depicted in
Fig. \ref{fig-model}. The presence of the chaotic cavity induces definite
phase changes (dephasing) in the resulting wave functions. Then, our main goal
will be to gain insight into how this dephasing results in the emergence of
decoherence. 

In what follows we will explore the effect of the coupled cavity on the phase
and the transmission probability through the system. Furthermore, we will also
address the consistency with B\"{u}ttiker's model of decoherence
\cite{cit-Buttiker-voltmeter} where the sample is coupled via a fictitious
voltage probe with a reservoir whose chemical potential is set to account for
current conservation. The presence of such reservoir accounts for a decoherent
\cite{cit-Baranger-Mello} re-injection of particles. A \textit{Hamiltonian}
formulation for this picture was proposed by D'Amato and Pastawski
\cite{cit-DAmato}. In that work, the connection of the dot states to an
infinite system with a continuous spectrum leads to a self-energy with an
imaginary part. This procedure is justified by considering decoherent electron
reservoirs within the Keldysh formulation \cite{cit-GLBE2,cit-Datta}. Here, we
re-examine the latter path by exploring the consequences of the coupling with
a system that contains a \textit{finite} number of states. From this point of
view, our main goal is to show how \textquotedblleft{decoherent}%
\textquotedblright\ behavior is an emergent phenomenon as the number of states
in the chaotic cavity increases. While in our discussion we adopt a single particle description,
the conclusions will be of a general nature \cite{cit-ChemPhys-e-ph}.

The total Hamiltonian is split into four terms
\[
\mathcal{H}=\mathcal{H}_{\mathrm{dot}}+\mathcal{H}_{\mathrm{electrodes}
}+\mathcal{H}_{\text{\textrm{cavity}}}+\mathcal{H}_{\mathrm{int.}}
\]
The device is represented by a Hamiltonian $\mathcal{H}_{\mathrm{dot}
}+\mathcal{H}_{\mathrm{electrodes}}$ consisting of a quantum dot that is
coupled through potential barriers to the left and right electrodes. In
addition, we introduce a chaotic cavity (represented by $\mathcal{H}
_{\text{\textrm{cavity}}}$) that serves as an \textquotedblleft
environment\textquotedblright\ that perturbs the system through the coupling
term contained in $\mathcal{H}_{\text{\textrm{int}}}$.

The dot sustains a set of states $\ E_{i}^{{}}$ whose corresponding creation
and annihilation operators are $d_{i}^{+}$ and $d_{i}^{{}}$ respectively. This
part of the Hamiltonian is written as
\begin{equation}
\mathcal{H}_{\text{\textrm{dot}}}=\sum_{i}E_{i}^{{}}d_{i}^{+}d_{i}^{{}}.
\end{equation}
The Hamiltonian for the electrodes is:
\[
\mathcal{H}_{\mathrm{electrodes}}=\sum_{k,\alpha=\mathrm{L},\mathrm{R}
}\varepsilon_{k\alpha}^{{}}c_{k\alpha}^{+}c_{k\alpha}^{{}}+\sum_{k,\alpha
}(V_{k\alpha,0}^{{}}c_{ks}^{+}d_{0}^{{}}+c.c.)
\]
where $c_{k\alpha}^{+}$ represents the creation in the eigenstate $k$ of the
electrode $\alpha$. The electron creation operators in an arbitrary basis for
the chaotic cavity are denoted by $b_{s}^{+}$
\begin{equation}
\mathcal{H}_{\text{\textrm{cavity}}}=\sum_{s=1}^{N}\varepsilon_{s}^{{}}
b_{s}^{+}b_{s}^{{}}+\sum_{s,r}^{N}(V_{s,r}^{{}}b_{s}^{+}b_{r}^{{}}+c.c.).
\end{equation}
Without loss of generality, the coupling between the chaotic cavity and the dot
is restricted to one (local) state:
\[
\mathcal{H}_{\mathrm{int}}=V_{s}^{{}}(b_{1}^{+}d_{0}^{{}}+c.c.).
\]
In order to simplify the physics we focus our attention on a single resonance
of energy $E_{0}$ which, being the closest to the Fermi energy, is relevant
for transport. The matrix elements $V_{k\alpha,0}$ describe the coupling between the
electrodes and the dot.
\begin{figure}[ptb]
\onefigure[width=12cm]{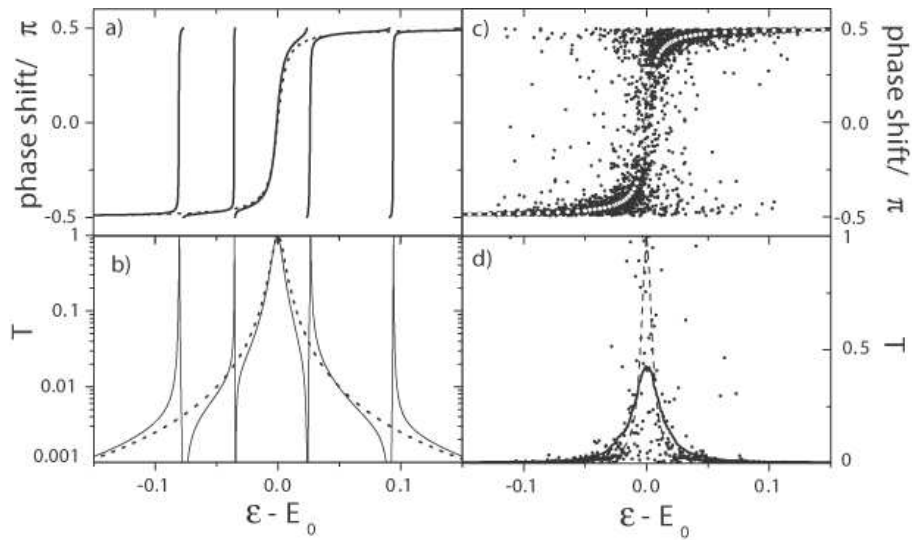}
\caption{ a) The phase shift as a
function of the electron energy. The solid line corresponds to the total
transmittance calculated using the model introduced in the text with only four
states in the chaotic cavity. The parameters of the Hamiltonian in units of
the hopping V in the leads are: E$_{0}=0,$ V$_{L}=$V$_{R}=0.05,$ V$_{s}=0.025$
and $\Delta_{N}=0.04$. For reference, the phase shift corresponding to the
situation of vanishing $V_{s}$ is shown with a dashed line. In b) we show the
total transmission probability as a function of the incident electron energy for the case discussed in panel a). In c) Scatter plot of the phase shift as a function of the incident electron
energy. The parameters of the Hamiltonian in units of the hopping V in the
leads are: E$_{0}=0,$ V$_{L}=$V$_{R}=0.05,$ V$_{s}=0.02$ and $\Delta_{N}=6.67\times
10^{-5}$. In these calculations 3000 states were included in the chaotic
cavity. For reference, the non-interacting phase shift is shown with circles
superposed to the scatter plot. In d) we show a scatter plot of the total
transmission probability. The solid line corresponds to the total
transmittance calculated using the coarse graining procedure. As in the
previous figures, the dashed line corresponds to the case of zero coupling
with the chaotic cavity.}
\label{fig-4states}
\end{figure}
The matrix elements of $\mathcal{H}_{\mathrm{cavity}}$ are assumed to be
distributed according to random matrix theory for the time-reversal invariant
case (gaussian orthogonal ensemble, GOE). The electrodes will be modeled as
one-dimensional tight-binding chains with hopping $V$.  $V_{L}$, $V_{R}$ and
$V_{s}$ are regarded as small parameters  compared with the band width of the electrodes.
The electrodes can be eliminated and their effect included exactly through a
self-energy ($^{L}\Sigma,$ $^{R}\Sigma$) \cite{cit-DAmato}. After
diagonalization of the matrix corresponding to $\mathcal{H}_{\mathrm{cavity}}
$, a similar procedure \cite{cit-Levstein-decim} can be applied for the \textquotedblleft finite
environment\textquotedblright\ thus obtaining a
self-energy $^{\mathrm{cavity}}\Sigma$. Note that in contrast to the
self-energy accounting for the electrodes that contain an imaginary part, here
$\operatorname{Im}(^{\mathrm{cavity}}\Sigma)=0$.

Once the self-energies are obtained, the calculation of the transmission
amplitude can be carried out by computing the retarded Green's function and
the group velocities at the electrodes \cite{cit-FisherLee}. It can be written
in terms of the decay rates \cite{cit-DAmato} as
\begin{equation}
t_{\mathrm{R,L}}=\mathrm{i}\hbar\sqrt{2^{\mathrm{R}}\Gamma}G_{\mathrm{R,L}
}^{R}\sqrt{2^{\mathrm{L}}\Gamma}\label{eq-transmission-amplitude},
\end{equation}
where $^{\mathrm{L}(\mathrm{R})}\Gamma=\operatorname{Im}(^{\mathrm{L}%
(\mathrm{R})}\Sigma)$.

In order to illustrate the basic physics of phase fluctuations, we consider
the simplified case in which there are only four levels in the chaotic cavity
and $\mathcal{H}_{\mathrm{cavity}}$ corresponds to a tridiagonal
matrix. In Fig. \ref{fig-4states} b) we show the transmission probability as a
function of the energy of the incident electrons. There we can appreciate
that, together with the main resonance, due to the state in the dot, there are
other resonances associated with the states in the cavity. The height of these
maxima is always one since the escape rates to the right and left electrodes
are equal \cite{cit-DAmatoAB}. Note that the width of these resonances is
decreased in comparison with the main resonance width because of the small
coupling $V_{s}$. We also emphasize the presence of \textit{antiresonances}
\cite{cit-Levstein-decim,cit-DAmatoAB} (i.e. zero-transmission points) in
this log-plot. The occurrence of these antiresonances is due to a
\textit{destructive interference} between the different possible
\textquotedblleft paths\textquotedblright\ connecting the left and right
electrodes. Such paths can be classified essentially as a direct path from
left to right, and paths that go from the left to the right electrode passing
through the chaotic cavity.

Another quantity of interest is the transmission phase \cite{cit-Hackenbroich}. This quantity is accessed experimentally   \cite{cit-phase-exp} by embedding a quantum dot in a branch of an
Aharonov-Bohm interferometer. From Eq. (\ref{eq-transmission-amplitude}) the
phase shift is given by
\begin{equation}
\theta(\varepsilon)=\frac{1}{2\mathrm{i}}\ln\left(  \frac{G_{\mathrm{R,L}}
^{R}(\varepsilon)}{G_{\mathrm{L,R}}^{A}(\varepsilon)}\right)
,\label{eq-phase-shift}
\end{equation}
where $G^{A}$ is the advanced Green's function \cite{phaseshift-note}. The
phase shift as a function of the electron energy is shown in Fig.
\ref{fig-4states}a. There, we can appreciate that at each resonance, the
phase shift experiences a smooth increase of $\pi$ through an energy of the
order of the resonance width. On the other hand, at each antiresonance the
phase-shift displays an \textit{abrupt} fall of $\pi$. This can be understood
by analyzing the path followed by the transmission amplitude in the complex
plane \cite{cit-HWLee}.

We study the case where $\mathcal{H}_{\mathrm{cavity}}$ has a dense spectrum
(characterized by a level spacing $\Delta_{N}\varpropto N^{-1})$ in the energy
region close to the main resonance. The phase shift as a function of the
electron energy is shown as a scatter plot in Fig. \ref{fig-4states}c. The 
phase shift for the case of a vanishing interaction with the chaotic
cavity is also shown for reference as empty circles.
The same behavior observed in Fig.\ref{fig-4states}a (rapid jumps of the phase) 
is now present but within a much smaller energy scale that is smaller than the
energy resolution of the plot. For this reason the phase values appear as scattered
points that give the phase at the particular energy values. If we had used a much finer
resolution in energy one would see the very fine structure of the antiresonances for the
dense spectrum. Due to the presence of antiresonances the phase shift is free to move between $-\pi/2$ and $\pi/2$. Hence, a small change in the energy of the
incoming electron can give rise to an important change in its phase. Here the
crucial point is to notice the energy is only resolved within a precision $\delta\varepsilon$
limited by the voltage bias and/or the thermal energy scale. In this range, antiresonances
produce uncontrolled fluctuations of the phase. This is what is understood
as dephasing, the practical impossibility to control and predict the phase of
a quantum state. Seen through a transport observable this results in decoherence.

Figure \ref{fig-4states}d shows a scatter plot of the transmission probability
as a function of the electron energy for 1000 energy values. For reference, the dashed curve
corresponds to the situation where there is no coupling with the chaotic
cavity. When coupling to the cavity is turned on, the original
resonance spreads in a bundle of thinner resonances not resolved in the plot.
If the spectrum of the cavity is dense enough (i.e., $\Delta_{N}
/\delta\varepsilon\ll1$ where $\delta\varepsilon$ is a given energy
resolution), the precise shape of these resonances is not only difficult to
determine numerically but clearly irrelevant! We then resort to a coarse
graining procedure accounting for the natural energy resolution on the energy scale $\delta\varepsilon$, and
compute the coarse-grained transmittance according to the prescription
\begin{equation}
\overline{T}_{\mathrm{R},\mathrm{L}}(\varepsilon)=\frac{1}{\delta\varepsilon}%
\int_{\varepsilon-\delta\varepsilon/2}^{\varepsilon+\delta\varepsilon
/2}T_{\mathrm{R},\mathrm{L}}(\varepsilon^{\prime})\mathrm{d}\varepsilon
^{\prime}.
\end{equation}
In the limit of large $N$ and small $\delta\varepsilon\gg\Delta_{N}$ we
recover the result of reference \cite{cit-DAmato}. In Fig. \ref{fig-4states}d, 
we show, with a solid line, the transmittance obtained using the coarse
graining procedure while the dashed line corresponds to the case $V_s=0$. For the
decoherent model \cite{cit-GLBE2} the transmittance, $\widetilde{T}_{\mathrm{R}
,\mathrm{L}}(\varepsilon)$, is the sum of a coherent and a decoherent term
 as
\begin{equation}
\widetilde{T}_{\mathrm{R},\mathrm{L}}(\varepsilon )=\frac{4~^{\mathrm{R}%
}\Gamma ~^{\mathrm{L}}\Gamma }{\left( \varepsilon -E_{0}\right) ^{2}+(^{%
\mathrm{L}}\Gamma +\,^{\mathrm{R}}\Gamma +\,^{\phi }\Gamma )^{2}}\,\left\{ 1+%
\frac{^{\phi }\Gamma }{^{\mathrm{L}}\Gamma +\,^{\mathrm{R}}\Gamma }\right\} .
\end{equation}
The numerical result for $\overline{T}_{\mathrm{R},\mathrm{L}}(\varepsilon)$
coincides with $\widetilde{T}_{\mathrm{R},\mathrm{L}}(\varepsilon)$ hence
verifying that
\begin{equation}\label{TRL}
\overline{T}_{\mathrm{R},\mathrm{L}}(\varepsilon)\underset{\delta
\varepsilon\rightarrow0}{\longrightarrow}\widetilde{T}_{\mathrm{R},\mathrm{L}%
}(\varepsilon).
\end{equation}
The conclusion is then that the coarse graining prescription applied to the coherent
transmittance returns the same two components as the classical application of
a voltage probe to the quantum dot. Note that $^{\phi}\Gamma$ is determined by the Hamiltonian parameters ($V_{s},\Delta_{N},N$) that can be estimated through a Fermi Golden Rule. We
performed similar calculations for a variety of models including several
energy levels and disorder in the dot consistently reproducing the results of
the voltage probe models. Since these equations provide a smooth crossover
from fully quantum coherent to classical incoherent transport \cite{cit-GLBE1},
phase fluctuations due to {\it antiresonances emerge as a road to decoherence}.

Similar results were obtained for a variety of models of the central system
coupled to single or multiple side cavities, each representing a dephasing channel, that can be addressed within the Landauer-B\"{u}ttiker approach. This confirms the generality of Eq.(\ref{TRL}). Almost any model of the side cavity suffices,
regardless of its chaoticity, as long as the density of states connected to
the central dot is dense. The particularities of the coupling are also unimportant (in our case to a single cavity level) since this will only shift the antiresonance positions and no qualitative changes in the spectra occur. The purpose of using a random matrix as a model for
the chaotic cavity is simply to avoid the appearance of particular spectral
regularities that could result in specific structures (Moir\'{e} patterns) in the discrete sampling of the transmittance.
In integrable systems, it might be practical to wash out possible patterns in
the level spacing using a Monte-Carlo procedure for the calculation of the
integral. Note also that in the Landauer-B\"{u}ttiker picture, the chemical
potential at the ficticious probe has to be determined in order to achieve the
voltmeter condition ($I_{\phi}=0$). Here, the finite size of the chaotic
cavity (implying $\operatorname{Im}(\Sigma_{\mathrm{cavity}})=0$ ) ensures the
conservation of the steady state current. In contrast, for a voltage probe the
corresponding self-energy has a non vanishing imaginary part. The equivalence
between both models is guaranteed by the complex structure of the real part of
$\Sigma_{\mathrm{cavity}}$ and the application of a coarse graining procedure.

The above results are clearly valid for a wide range of physical situations where
the essential phenomenon is wave propagation (i.e. electrons \cite{cit--Webb-Washburn,Kobayashi}, spins \cite{Madi} or electromagnetic fields \cite{Rotters}). 
A hint on how the above arguments play in a full many-body problem already
appears when we consider the interaction with a phonon mode. In this solvable
case, one can appreciate that decoherence arises \cite{cit-Bonca-97} because
the interaction with the phonons opens several orthogonal outgoing channels
for the electrons, each one differing in the number of phonons in the system
\cite{cit-ChemPhys-e-ph,cit-Makler}. In this sense, our model can be thought of
as a special kind of electron-phonon model where only virtual phonon emission or
absorption is allowed.

In Ref. \cite{cit-Imry-phase-breaking} Stern, Aharonov and Imry argued that
decoherence can be explained in terms of either a change induced by the
particle in the environment that shifts it to an orthogonal state, or by a
randomization of the particle's phase. In our case, we can clearly see that
the coupling with the cavity is manifested in the dephasing. Besides, the idea
of the bath as a generator of decoherence arises here from the impossibility
to control the phase of the wave function and by restricting the energy
resolution when evaluating the coarse-grained transmittance. Indeed, the
maximum between the thermal uncertainty $k_{B}T$ and the applied voltage
provides a natural limit for the energy resolution. Experimentally, this means
that if this system is placed in an arm of an Aharonov-Bohm interferometer,
the crossover from the situation $\Delta_{N}/\delta\varepsilon\ll1$ to
$\Delta_{N}/\delta\varepsilon\gg1$ should show up as an attenuation of the
amplitude of the interferences.

We acknowledge financial support from ANPCyT, SeCyT-UNC, CONICET (Argentina)
and FONACIT (Venezuela). We thank the Abdus Salam ICTP (Trieste, Italy) for
the hospitality.


\begin{thebibliography}{99}
\bibitem {phaseshift-note}Notice also that the phase shift is related with the
Wigner time delay $\tau_{W}=\hbar\dfrac{\partial\theta(\varepsilon)}%
{\partial\varepsilon}$. Thus, it is also important from the point of view of
the transport dynamics Ref.\cite{cit-GLBE2}.
\end{thebibliography}
\end{document}